\documentclass[prl,twocolumn,showpacs,preprintnumbers,amsmath,amssymb]{revtex4}

\usepackage{graphicx}
\usepackage{epsfig}
\usepackage{dcolumn}
\usepackage{bm}

\begin{document}

\title{Phase Dynamics of Ferromagnetic Josephson Junctions}

\author{I. Petkovi\'{c}$^{*}$ and M. Aprili}
\affiliation {Laboratoire de Physique des Solides, UMR 8502, B\^{a}t.$\,
 \, 510$,~Universit\'{e} Paris-Sud, 91405 Orsay Cedex, France.}

\altaffiliation {petkovic@lps.u-psud.fr}

\date{\today}

\begin{abstract}

We have investigated the classical phase dynamics of underdamped ferromagnetic Josephson junctions by measuring the switching probability both in the stationary and non-stationary regime down to 350 mK. We found the escape temperature to be the bath temperature, with no evidence of additional spin noise. In the non-stationary regime, we have performed a pump-probe experiment on the Josephson phase by increasing the frequency of the junction current bias. We show that an incomplete energy relaxation leads to dynamical phase bifurcation. Bifurcation manifests itself as premature switching, resulting in a bimodal switching distribution. We directly measure the phase relaxation time $\tau_\varphi$ by following the evolution of the bimodal switching distribution when varying the bias frequency. Numerical simulations account for the experimental values of $\tau_\varphi$.

\end{abstract}


\maketitle

The maximum dissipationless current that a Josephson junction can carry, $I_s$, is not only given by the coupling energy between the two superconductors forming the junction, but also by the dynamics of their phase difference $\varphi$ \cite{tinkham_book}. Thermal and quantum fluctuations of the phase, for instance, lower $I_s$ from the critical current value $I_C=\pi \Delta/2 e R_n$, where $\Delta$ is the superconducting gap and $R_n$ the junction resistance in the normal state \cite{baratoff}. Furthermore, they introduce in a current-biased junction a probability distribution of the threshold current to switch into the dissipative state \cite{webb}.
Two time scales are pertinent: the inverse plasma frequency, $\omega_0^{-1}$, and the phase relaxation time, $\tau_\varphi$, which is defined by the junction's damping \cite{tinkham_book}. Since the plasma frequency is in the gigahertz range, it is actually the microwave impedance, including that of the junction environment, that sets the relaxation. However, in large-area tunnel junctions, the effect of the environment is negligible as the quasiparticle resistance is smaller than the vacuum impedance and the phase dynamics is "intrinsic" in the sense that it only depends on the junction parameters \cite{martinis}.

In this Letter, we address the classical phase dynamics of large-area underdamped ferromagnetic Josephson junctions. We show that in the stationary regime, i.e. when the current bias frequency $\omega_b$ is much smaller than the inverse phase relaxation time $\tau_\varphi^{-1}$, the mean switching current $\langle I_s \rangle$ and its standard deviation $\sigma$ are as expected for thermal fluctuations. Whereas in the non-stationary limit, i.e. when the bias frequency is comparable to the inverse phase relaxation time,
$\langle I_s \rangle$ has a finite probability to jump to zero, thus showing an apparent zero critical current. The premature switching reveals the phase bifurcation. The exponential rise of the premature switching probability $N_1$ as a function of the bias frequency $\omega_b$ gives the phase relaxation time $\tau_\varphi$ in agreement with the numerical simulations. Moreover, no effects of the magnetization on the phase dynamics at zero voltage are observed.

The phase dynamics of a Josephson junction is usually described within the Resistively Shunted Josephson junction (RSJ) model \cite{stewart} as

\vspace{-10pt}

\begin{equation}
\ddot{\varphi} +  \beta \dot{\varphi}+ \sin \varphi = \eta_b(t),
\label{rsj}
\end{equation}

\vspace{2pt}

\noindent where $\beta=(RC\omega_0)^{-1}$, $\omega_0=\sqrt{2 e I_C/\hbar C}$ is the plasma frequency, and $\eta_b=I_B/I_C$, where $I_B(t)$ is the time dependent bias current. In our experiment it is a sawtooth of frequency $\omega_b$. The time is normalized to $\tau = \omega_0 t$, and the dot denotes $d/d\tau$.
Here $C$ is the junction capacitance and $R$ the quasiparticle tunneling resistance.
Damping $\beta$ is set by the $RC$ product. Escape by thermal fluctuations has been largely investigated since the pioneering work of Fulton and Dunkelberger \cite{fulton}. Thermal escape is usually described using the Kramers theory
in the harmonic approximation \cite{kramers}. The kinetic energy of the order of the Josephson energy, and much higher than the thermal energy, makes the harmonic approximation incorrect. The anharmonicity produces bifurcation \cite{landau} in the phase dynamics that in the end strongly modifies the switching distribution probability, rendering it bimodal.

Bifurcation usually occurs when an anharmonic oscillator is driven above a certain critical amplitude close to the resonant frequency. This creates two stable attractors in the phase space
and hence two macroscopically different states. An example of a system in which bifurcation occurs is the Duffing oscillator \cite{nayfeh}. Dynamically driven bifurcation is often referred to as a kinetic phase transition \cite{dykman}. The noise allows for stochastic transitions between the two attractors leading to the finite probability of finding the system in each state.
Microwave-induced bifurcation for the Josephson junctions has been observed in the stationary regime \cite{jensen} and  has been realized experimentally in  the pioneering work by Yurke et al. \cite{yurke} for parametric amplification \cite{devoret,cleland}.

In the non-stationary regime \cite{barone}, it is the initial kinetic energy that induces the phase bifurcation.
Two phase states, shown in Fig.~1(a), are achieved: Either the phase relaxes to the bottom of the potential well resulting in a switching current $I_s$ equal to that in equilibrium, or the phase does not relax and the escape occurs for much smaller current bias, of the order of the retrapping current $I_r$. In strongly underdamped junctions as these, the retrapping current is vanishingly small.  The two solutions are illustrated in the Fig.~1(b), showing the numerically obtained $V(t)$ from the RSJ model [Eq.~(\ref{rsj})], considering a periodic bias current with a sawtooth waveform. After the first escape at $I_s$, the kinetic energy is large enough for premature escape at $I_r$ when the current is ramped up again. Thus in the non-stationary limit the kinetic energy induces a phase transition between the locked and the running state \cite{landauer}. This is illustrated in the calculated phase diagram $(\varphi,\dot{\varphi})$ [Fig.~1(c)]. The full and dotted curves correspond to the locked and running state respectively.  The initial kinetic energy, i.e., initial speed $\dot{\varphi}(t\!=\!0)$ at the start of each cycle, determines the phase trajectory, and the dashed line is (approximately) the separatrix between the running and the locked state trajectories.

\begin{figure}[h!]
\centerline{\hbox{\epsfig{figure=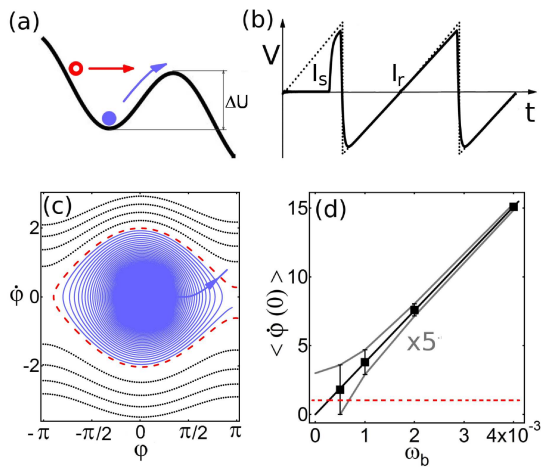,width=75mm}}}
  \caption{(color online). (a) The switching mechanism in the non-stationary regime, where the phase can either relax to the bottom of the  potential well and then escape (solid circle), or stay near the top and escape early (open circle). (b) The calculated voltage $V(t)$ across the junction (full curve), with the imposed current sawtooth ramp (dotted curve). Either the switching takes place as in equilibrium, around $I_s$, or around the retrapping current $I_r$.  (c) The phase diagram, calculated for $\beta=0.01 \omega_0$, $\omega_b=0.001\omega_0$ and $\eta_b=1.2$. The attractor (solid curve) corresponds to the locked state, while the dotted curves correspond to the running state. In the case of the early escape, the phase trajectory is near the separatrix (dashed curve). (d) Numerically obtained distribution of the initial phase speed at the start of each cycle $\langle \dot{\varphi}(0) \rangle$, as a function of the ramp frequency. The error bars are augmented 5 times. Full line is a linear fit, and the dashed line is the bifurcation threshold corresponding to the separatrix.
}
  \label{one}
\end{figure}

The
dynamical bifurcation depends on the bias frequency, bias amplitude and damping. It turns out that it occurs only when the phase relaxation time, $\tau_\varphi$, is comparable with the driving force period $2\pi/\omega_b$. Note that $\tau_\varphi$ in strongly underdamped junctions is 4 orders of magnitude larger than the resonant period $2\pi/\omega_0$. Therefore, the out-of-equilibrium conditions shift the time scale for bifurcation from the resonant frequency, as expected in the stationary regime, down to the phase relaxation time. The numerical solution of the RSJ model shows that no premature switching is expected when $\tau_\varphi \ll 2 \pi/\omega_b$, while it is indeed observed at $\tau_\varphi \sim 2 \pi/\omega_b$. For $\tau_\varphi \gg 2 \pi/\omega_b$ we find only the premature switching solution. Physically, in this case the phase has not relaxed when a small driving force (i.e., a current bias equal to the retrapping current) allows it to escape.
The effect of noise in the numerical model is replaced by the fluctuations of the initial kinetic energy for each current sweep due to an incomplete relaxation. These fluctuations allow for stochastic transitions between the locked and the running state. The numerically obtained initial speed (averaged over many periods) $\langle \dot{\varphi}(0) \rangle$ as a function of the ramp frequency in Fig.~1(d) sums up the effect of the bias frequency on the kinetic phase transition. The fluctuations of the initial speed are given by the error bars. The dashed line is the separatrix, so when the initial speed is over the bar, there is a finite probability for bifurcation.

In the nonstationary limit, the probability for premature switching  $N_1$, obtained by numerically solving  the RSJ model \cite{thesis} [Eq.~(\ref{rsj})] for a sawtooth ramp of amplitude $\eta_b$ and frequency $\omega_b$,  is well described by

\vspace{-10pt}

\begin{equation}
N_1\!=\!1\!-\!A \exp (- \tau_\varphi \omega_b)\!=\!1\!-\!1.8 \exp\! \left[- \,0.76\, \frac {\eta_b\, \omega_b}{\beta^{\frac{3}{2}}} \right]\!.
\label{emp_formula1}
\end{equation}

\vspace{2pt}

\noindent At $\tau_\varphi \!\sim \!2\pi/\omega_b$, the current switching distribution is bimodal, and $N_1$ is the surface of the histogram at $I_r$. In a pump-probe experiment, the relaxation time is usually obtained by tracking the time needed to reach equilibrium after an external perturbation is turned off. Here, similarly, the phase relaxation time can be directly probed by measuring the early switching probability $N_1$ as a function of the bias frequency. This offers an alternative way to measure oscillation damping \cite{clarke}.

\textit{Experimental.} We fabricate large-area superconduc\-tor-\-insulator-\-ferromagnet-superconductor (SIFS) tunnel junc\-tions  by electron gun evaporation through shadow masks in ultrahigh vacuum. The superconductor is Nb, the insulator is $\rm Al_2O_3$,  the ferromagnet a PdNi alloy with 10 $\%$ Ni, and junction surface is $0.6 \times 0.8$ $\rm mm^2$. The fabrication technique is described elsewhere \cite{kontos}. In this Letter, we investigate escape in Josephson junctions showing only $\pi$ coupling and corresponding to a thickness of the PdNi ferromagnetic layer $d_F$ varying from 70 to 100 $\AA$. We concentrate on the $\pi$ junction as it has been used as a phase battery in superconducting circuits \cite{dellarocca,bauer}, and it has been argued that it may be used in superconducting $Q$-bits \cite{ioffe}. The coupling between the phase and the spin dynamics of the PdNi thin layer is an important issue. Furthermore, from a technical point of view, the ferromagnetic layer suppressed the critical current by a factor of 100, reducing substantially the power dissipated upon switching, keeping the heating negligible.

Junctions have been measured in a $\rm ^3 He$ cryostat from 4.2 K down to 350 mK. The current-voltage curve of a typical SIFS junction at 350 mK is presented in Fig.~2(a). The curve is that expected for a strongly underdamped junction. The critical current is $I_C\!=\!130$ $\mu A$. The junction resistance in the normal state is $R_n\!=\! 0.2 \,\Omega$, and capacitance is $C\!=\!42$ nF, as estimated from the $\rm Al_2O_3$ dielectric constant and the Fiske steps. We obtain the plasma frequency $\omega_0\!=\!3 \times 10^9$ rad/s, and the quality factor $Q\!=\!\beta^{-1}\! =\! 3000$. We estimate the retrapping current to be $I_r\!\sim\! 4 I_C/\pi Q \!\sim \!50$ nA. Finally, the power dissipated at switching is of the order of 1 nW. As the junction volume is about $10^5$ $\mu \rm m^3$, the cooling power from the electron-phonon coupling leads to an irrelevant increase of the electron bath temperature.

The switching current distribution is measured using a counter. The junction is current-biased with a sawtooth waveform, and the switching current is easily obtained by multiplying the time before dissipation and the speed of the ramp. Positive and negative bias give the same result. All of the incoming and outgoing lines are filtered with band pass filters with the cutoff frequency of 1 MHz, at 4.2 K and room temperature.

\begin{figure}[h!!!]
\centerline{\hbox{\epsfig{figure=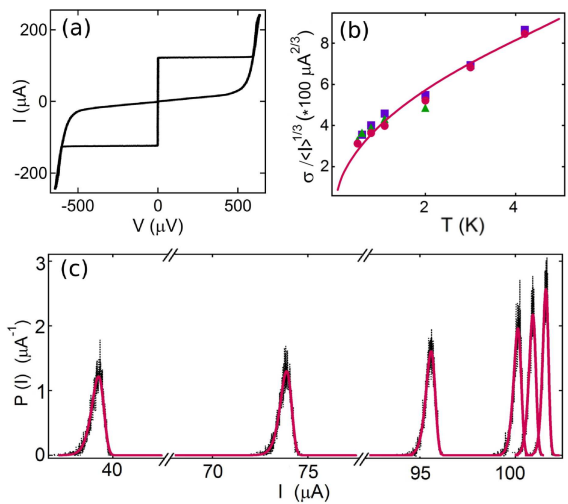,width=75mm}}}
  \caption{(color online). (a) Typical $IV$ curve at 350 mK. (b) Distribution width as a function of temperature for junctions with 70, 80, and 85 $\AA$ of PdNi (triangles, circles, and squares, respectively). The solid curve is a fit for the junction with 80~$\AA$ in the low damping regime. (c) The measured histograms $P(I)$ (dotted curves) for $T=$ 0.5, 0.8, 1.1, 2.0, 3.0, and 4.2 K, respectively, from the right, and a fit (solid curves), for a junction with 80 $\AA$ of PdNi. The effective temperature, obtained as a fitting parameter, is almost equal to the bath temperature.
}
  \label{two}
\end{figure}

\textit{Thermal Escape.}
The switching current distribution of the junction with 80 $\AA$ of PdNi, measured at different temperatures, is presented in Fig.~2(c) for the bias frequency of 37 Hz. This time scale is much longer than the phase relaxation time, and hence the phase follows adiabatically the drive. The mean switching current increases linearly with decreasing temperature \cite{kontos}. The width narrows as expected from escape induced by thermal fluctuations.
The fit (solid curves) is obtained using the Kramers theory \cite{fulton,kurkijarvi} for  low damping.

The distribution width $\sigma$ as a function of bath temperature $T$ is presented in Fig.~2(b) for three different junctions, with $d_F\!=\!70$, 80 and 85 $\AA$ of PdNi.
The solid curve is the fit in the low damping regime, which
takes into account the temperature-dependent quasiparticle resistance $R$ \cite{ruggiero}.
No phase noise induced by spin noise is observed. Note that here we focus on the classical phase dynamics, and no crossover to the quantum tunneling is expected down to 10 mK.
Spin noise originating from thermal fluctuations of the magnetization can affect the escape through the fluctuations of the critical current $\delta I_M$ or through the magnetization-induced phase noise $\delta \varphi_M$. The critical current depends on the exchange energy as $I_C\!\!\sim \!\exp(-d_F /\xi_F)\cos(d_F /\xi_F)$ \cite{buzdin}, with $\xi_F\!=\!\sqrt{\hbar D/E_{ex}}$, where $E_{ex}$ is the exchange energy and $D$ the diffusion coefficient. For  $D\!\sim\!10\,\rm cm^2/s$, $E_{ex}\!\sim\!25$ mV and $d_F\!\sim\!80\,\AA$, we have $d_F/\xi_F\!\sim\!\pi$ and $\delta I_M/I_C$ goes as $(\delta E_{ex}/E_{ex})^2$. Therefore, for $\delta E_{ex}\!\sim\!k_B T$, we get $\delta I_M\!\sim\!10$ nA, which is below our resolution.
The overall $\delta \varphi_M$ produced by thermal domain wall motion and intradomain oscillations of the magnetization is small in large-area junctions. Thermal phase fluctuations account well for the switching current distribution measured at equilibrium.

\textit{Phase Relaxation Time.}
In order to investigate the change in the switching current distribution when the phase is nonstationary, we systematically increase the bias frequency. We observe that current switching distribution changes drastically and becomes bimodal above 5 kHz. With increasing bias frequency, the number of counts around $I_s$ decreases, while the number of counts around $I_r$ increases. To increase sensitivity, instead of the retrapping current $I_r$, we measured the current corresponding to the first Fiske resonance $I_f \!\sim\! 15\, \mu$A \cite{fiske}. This is shown in Fig.~3(a), where we report the switching current distribution for three different bias frequencies. The total number of sweeps is 2000.
Above 20 kHz,
the mode around $I_s$ disappears.
The histograms at high bias frequency show only a tiny reduction in the mean value of $I_s$ with respect to that measured at equilibrium, while almost no change in their width indicates negligible thermal effects.
It is worthwhile to mention that we have observed this behavior in Josephson junctions without any ferromagnetic layer when the critical current is almost canceled by applying a quantum of magnetic flux in the junction (i.e., around the first minima in the Fraunhofer pattern).


\begin{figure}[t!!!]
\centerline{\hbox{\epsfig{figure=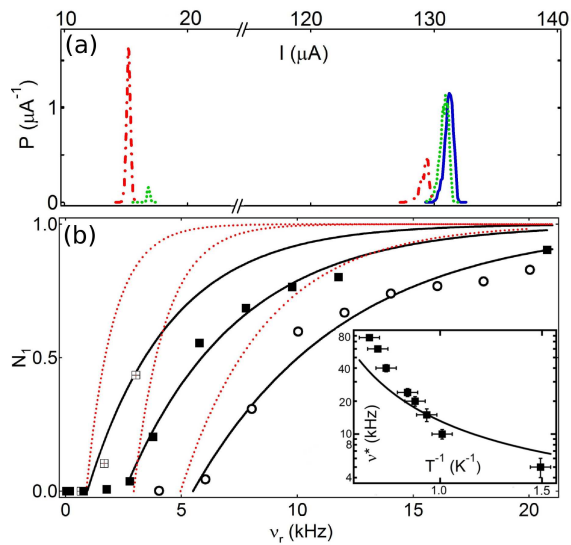,width=75mm}}}
  \caption{(color online). (a) Switching histogram of the junction with 85 $\AA$ of PdNi taken at 350 mK with the sweep frequency of 4 kHz (solid curve), 6 kHz (dotted curve) and 12~kHz (dashed-dotted curve). (b) Early switching probability $N_1$ as a function of ramp frequency, measured at 350 mK for the junctions with 85 (open circles), 100 (solid squares) and 90 $\AA$ (crossed squares) of PdNi. The dotted lines are a theoretical prediction [Eq.~(2)]. The solid curves are obtained from the same equation, when keeping $\tau_\varphi$ as the fitting parameter. Inset: The dependence of the bifurcation onset frequency $\nu^*$ of temperature, measured on a junction with 85~$\AA$. The solid line is a fit taking into account the measured quasiparticle resistance $R$.
}
  \label{three}
\end{figure}

The premature switching probability $N_1$ as a function of bias frequency is presented in Fig.~3(b).
The dotted curves are the solutions of the RSJ model [Eq.~(2)] without fitting parameters,
while the solid curves are the best fit, giving $\tau_\varphi=157, \,207,$ and $282\,\,\mu $s, respectively, for junctions with 85, 100, and 90 $\AA$ of PdNi \cite{remark2}.
The discrepancy could arise from the fact that in the numerical model we do not account for thermal noise. The frequency at which the bifurcation starts, $\nu^*$, is reproduced numerically.
An independent way to measure the $RC$ product is to measure the delay of $V(t)$ relative to the start of the ramp, and the obtained values are in accord with the experimental $\tau_\varphi$.


Finally, we show the measured bias frequency at which bifurcation starts, $\nu^*$, as a function of temperature [inset in Fig.~3(b)]. The fit is obtained numerically as a solution of $N_1(\nu^*)\!=\!0$, taking into account the measured quasiparticle resistance.

\textit{Conclusions.} We measure the switching histograms of the strongly underdamped SIFS Josephson junctions. In the stationary regime, the escape temperature is the bath temperature indicating negligible magnetic-induced noise. Therefore at least the large-area $\pi$ junction seems to provide a noiseless phase battery. In the nonstationary regime, when the junction is ramped before complete phase relaxation, we observed a kinetic phase transition induced by phase bifurcation.  The probability of measuring a finite switching current decays as a function of the bias frequency, allowing to measure the phase relaxation time as in a pump-probe experiment. From a practical point of view, dissipationless current of a Josephson junction can be the turned on and off by changing the bias frequency.

We thank J. Gabelli, B. Reulet, T. Kontos and Z. Radovi\'{c} for stimulating discussions.


\end{document}